\documentclass[runningheads,openany]{svmult}

\usepackage{makeidx}   % allows index generation
\usepackage{graphicx}  % standard LaTeX graphics tool
\usepackage{subeqnar}  % subnumbers individual equations
\usepackage{multicol}  % used for the two-column index
\usepackage{physprbb}  % modified textarea for proceedings,
\usepackage{epsfig}
\usepackage{latexsym}

\newcommand{\klesssim}{\mathrel{\hbox{\rlap{\hbox{\lower4pt\hbox{$\sim$}}}\hbox{$<$}}}}

\newcommand{\llsim}{\mathrel{\lower4pt\hbox{$\sim$}}\hskip-11.5pt\raise1.6pt\hbox{$<$}\;}

\begin{document}

%%%%%%%%%%%%%%%%%%%%%%%%%%%%%%%%%%%%%%%%%%%%%%%%%%%%%%%%%%%%%
%%%%%%%%%%%%%%%%%%%%%%%%%%

\mainmatter
\label{03}

\title*{Millisecond Pulsars\protect\newline as Tools of Fundamental Physics}
\toctitle{Millisecond Pulsars as Tools of Fundamental Physics}
% allows explicit linebreak for the table of content
%
%
\titlerunning{Millisecond Pulsars}
% allows abbreviation of title, if the full title is too long
% to fit in the running head
%
\author{Michael Kramer}
\authorrunning{Michael Kramer}
% if there are more than two authors,
% please abbreviate author list for running head
%
%
\institute{University of Manchester, Jodrell Bank Observatory,
Cheshire SK11 9DL, UK}

\maketitle              % typesets the title of the contribution

\begin{abstract}
A new era in fundamental physics began when pulsars were discovered in
1967. Soon it became clear that pulsars were useful tools for a wide
variety of physical and astrophysical problems. Further applications
became possible with the discovery of the first binary pulsar in 1974
and the discovery of millisecond pulsars in 1982.  Ever since pulsars
have been used as precise cosmic clocks, taking us beyond the
weak-field limit of the solar-system in the study of theories of
gravity. Their contribution is crucial as no test can be considered to
be complete without probing the strong-field realm of gravitational
physics by finding and timing pulsars. This is particularly
highlighted by the discovery of the first double pulsar system in
2003. In this review, I will explain some of the most important
applications of millisecond pulsar clocks in the study of gravity and
fundamental constants.
\end{abstract}

\section{Introduction}

The title of this volume, ``Astrophysics, Clocks, and Fundamental
Constants'', would also be a suitable title for this contribution
describing the use of radio pulsars in the study of fundamental
physics.  Indeed, pulsar astronomy is an extraordinary discipline
which removes the distinction between physics and astrophysics that is
often made. Such a distinction may be justified by the fact that in a
terrestrial laboratory we can modify the experimental
set-up and control the environment. In contrast, in astrophysical
experiments we remain an observer, deriving all our information simply
from observing photons and their properties. Thereby, terrestrial
experiments are typically more precise and, most importantly, can be
reproduced -- at least in principle -- in any other laboratory on
Earth. However, when probing the limits of our understanding of
fundamental physics, we often have to study conditions that are too
extreme to be encountered on Earth. One may take the experiment into
space, like ``LISA'', ``STEP'' or ``Gravity Probe-B'', but even then
we are limited, particularly if we want to study gravity. While solar
system tests provide a number of very stringent tests for general
relativity, none of the experiments made or proposed for the future
will ever be able to test the strong field limit. For such studies,
pulsars are and will remain the only way to test and enhance our
understanding.  Additionally, pulsars not only provide us with the
only means to perform strong-field experiments, but these experiments
are also amazingly precise. It is this unique aspect that I will
review in the following. The interested reader may also consult the
excellent reviews by Will \cite{03wil01} and Turyshev \cite{03Tur} on
PPN formalism, by Wex \cite{03wex01} and Stairs \cite{03sta03} on strong
gravity tests, and by Lorimer \cite{03lor01} on pulsars in general.

\section{Pulsars}

\begin{figure}[h]
\begin{minipage}{6cm}
\includegraphics[width=1.1\textwidth]{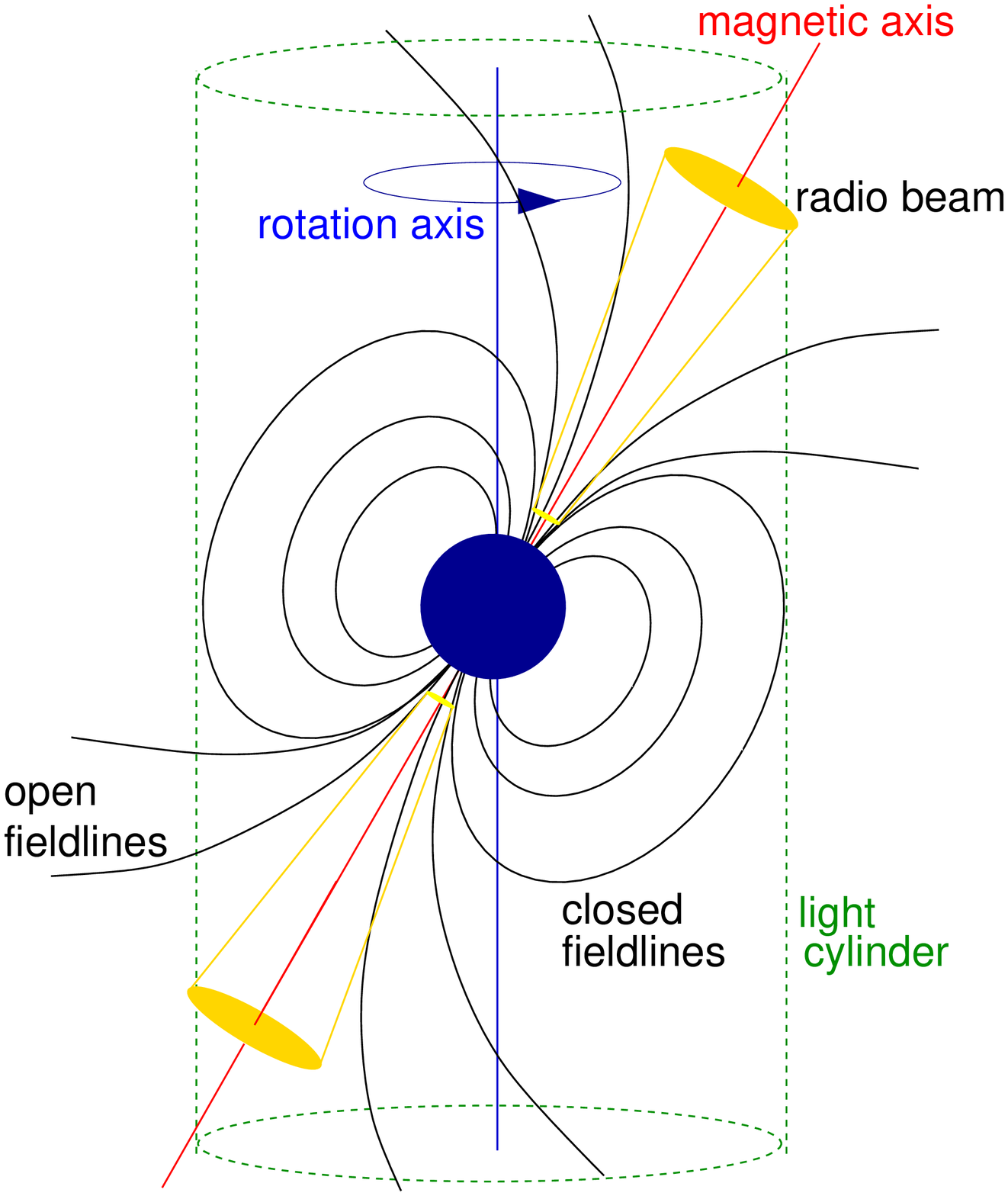}
\end{minipage}
\hspace{1cm}
\begin{minipage}{6cm}
\centerline{\includegraphics[width=.8\textwidth]{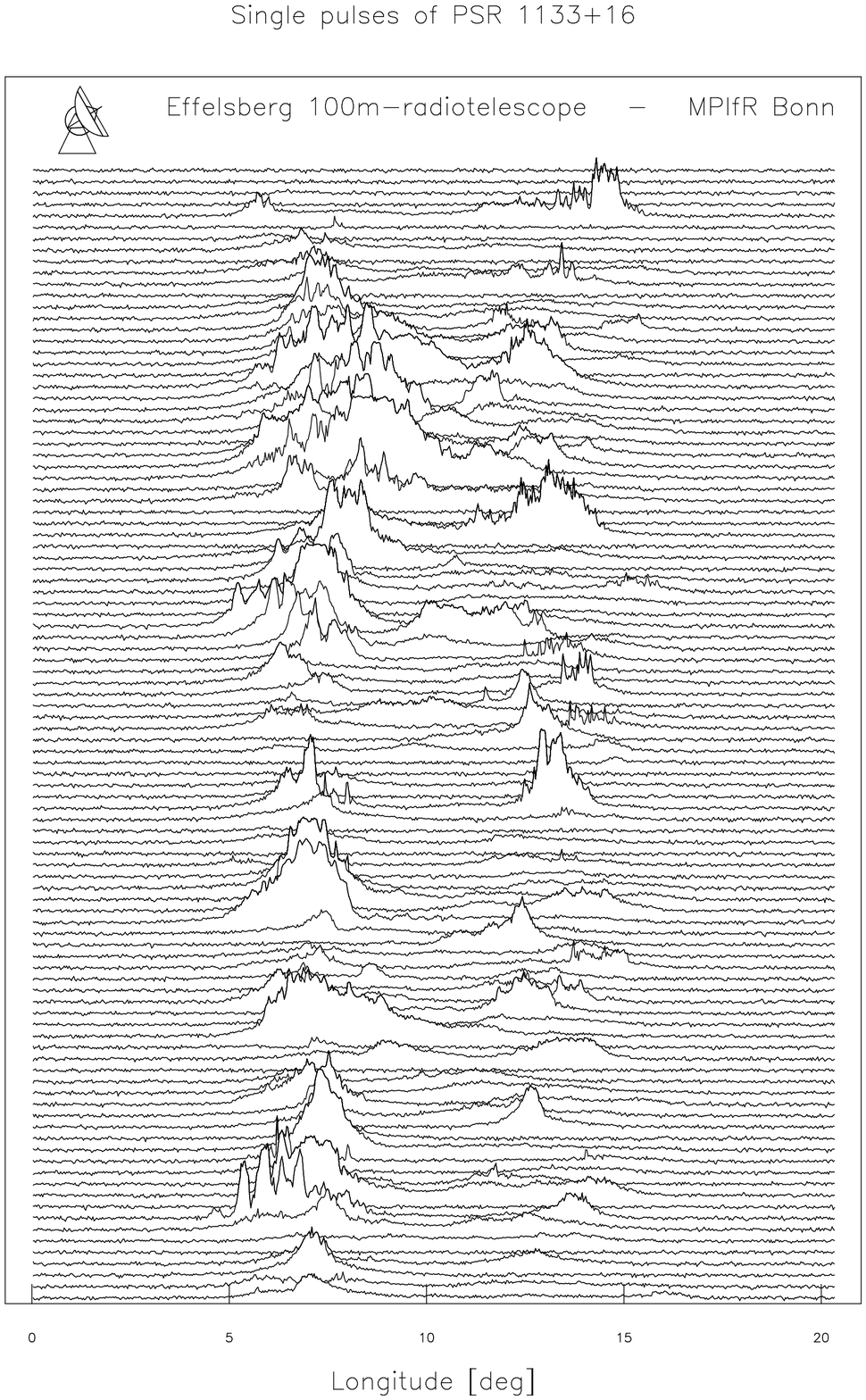}}
\centerline{\includegraphics[width=.7\textwidth]{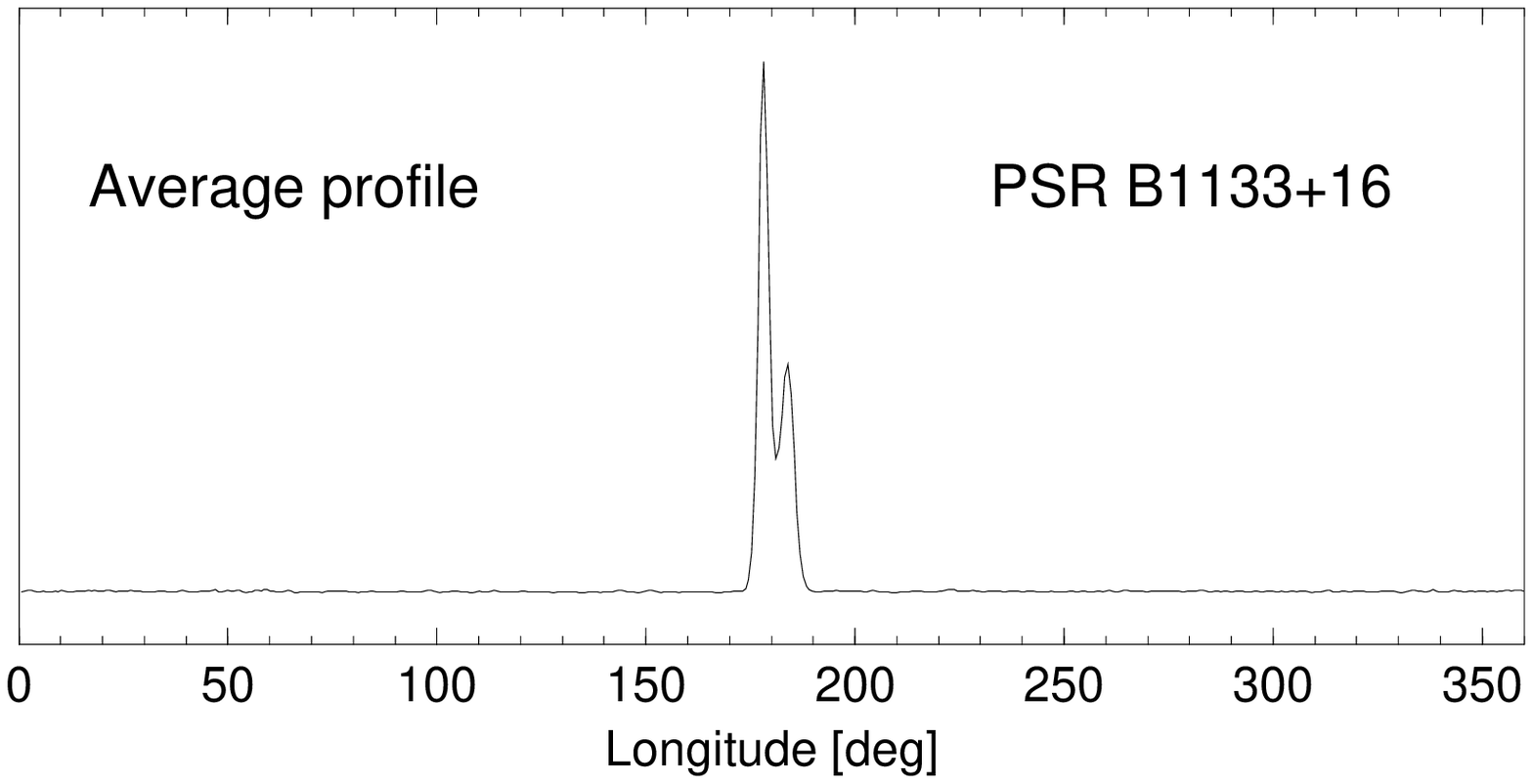}}
\end{minipage} 

\caption[]{(left)
A pulsar is a rotating, 
highly magnetised neutron star. A radio beam
centred on the magnetic axis is created at some height above the surface.
The tilt between the rotation and magnetic axes makes the pulsar 
in effect a cosmic lighthouse when the beam sweeps around in space.
(right) Individual pulses vary in shapes and strength (top)
average profiles are stable (bottom). The typical pulse width
is only $\sim$4\% of the period.
}
\label{03fig:pulsar}
\end{figure}

Pulsars are highly magnetized, rotating neutron stars which emit a
narrow radio beam along the magnetic dipole axis. As the magnetic
axis is inclined to the rotation axis, the pulsar acts like a cosmic
light-house emitting a radio pulse that can be detected once per
rotation period when the beam is directed towards Earth
(Fig.~\ref{03fig:pulsar}). For some very fast rotating pulsars, the
so-called millisecond pulsars, the stability of the pulse period is
similar to that achieved by the best terrestrial atomic clocks. This
is not surprising if we consider that they have large rotational
energies of $E=10^{43-45}$ J and low energy loss rates.  Using these
astrophysical clocks by accurately measuring the arrival times of
their pulses, a wide range of experiments is possible, some of which
are presented here. While it is not of utmost importance for the
remainder of this review {\em how} the radio pulses are actually
created, we will consider some of the basic pulsar properties below.

\subsection{Pulsars as Neutron Stars}

Pulsars are born in supernova explosions of massive stars.  Created in
the collapse of the stars' core, neutron stars are the most compact
objects next to black holes. From timing measurements of binary
pulsars (see Section~\ref{03dns}), we determine the masses of
pulsars to be within a narrow range of $(1.35\pm0.04)\,
M_\odot$\cite{03tc99}. Modern calculations for different equations of
state produce results for the size of a neutron star quite similar to
the very first calculations by Oppenheimer \& Volkov \cite{03ov39},
i.e.~about 20 km in diameter.  Such sizes are consistent with
independent estimates derived from modelling light-curves and
luminosities of pulsars observed in X-rays (e.g.~\cite{03zp98}). 

As rotating magnets, pulsars emit magnetic dipole radiation as
the dominant effect for an increase in rotation period, $P$,
described by $\dot{P}$.
Equating the corresponding
energy output of the dipole to the loss rate in rotational energy, we
obtain an estimate for the magnetic field strength at the
pulsar surface from
\begin{equation}
B_S=3.2\times 10^{19} \sqrt{P\dot{P}} \;\mbox{\rm Gauss},
\label{03fig:bfield}
\end{equation}
with $P$ measured in s and $\dot{P}$ in s s$^{-1}$.
Sometimes twice the value is quoted to reflect the field at the poles.
Typical values are of order $10^{12}$ G, although field strengths up
to $10^{14}$ have been observed \cite{03msk+03}.
Millisecond pulsars have lower field strengths of the order of $10^8$
to $10^{10}$ Gauss which appear to be a result of their evolutionary
history (see Section~\ref{03life}).  These magnetic
fields are consistent with values derived from X-ray  spectra
of neutron stars where we observe 
cyclotron lines \cite{03bclm03}.

\subsection{Pulsars as Radio Sources}

The radio signal of a pulsar is usually weak, both because the pulsar
is distant and the size of the actual emission region is
small. Estimates range down to a few metres, resulting in brightness
temperatures of up to $10^{37}$ K \cite{03hkwe03}. Such values require a
coherent emission mechanism which, despite 35 years of intensive
research, is still unidentified. However, we seem to have some basic
understanding, in which the magnetized rotating neutron star induces
an electric quadrupole field which is strong enough to pull out
charges from the stellar surface (the electrical force exceeds the
gravitational force by a factor of $\sim10^{12}$!). The magnetic field
forces the resulting dense plasma to co-rotate with the pulsar. This
{\em magnetosphere} can only extend up to a distance where
the co-rotation velocity reaches the speed of light\footnote{Strictly
speaking, the Alfv\'en velocity will determine the co-rotational
properties of the magnetosphere.}. This distance defines the so-called
light cylinder which separates the magnetic field lines into two
distinct groups, i.e.~{\em open and closed field lines}. The plasma
on the closed field lines is trapped and co-rotates with the pulsar
forever. In contrast, plasma on the open field lines can reach highly
relativistic velocities and can leave the magnetosphere, creating the
observed radio beam at a distance of a few tens to hundreds of km
above the pulsar surface (e.g.~\cite{03kxj+97}, 
see Fig.~\ref{03fig:pulsar}).

Most pulsars are not strong enough for us to allow studies of their
individual radio pulses. Then, only an integrated pulse shape, the
``pulse profile'', can be observed. Individual pulses reflect the
instantaneous plasma processes in the pulsar magnetosphere, resulting
in often seemingly random pulses (see Fig.~\ref{03fig:pulsar}). In
contrast, the average pulse profile reflects the global constraints
mostly given by a conal beam structure and geometrical factors and is
thereby stable. It is this profile stability which 
allows us to time pulsar to high precision.

\begin{figure}[h]
\centerline{\includegraphics[width=0.55\textwidth]{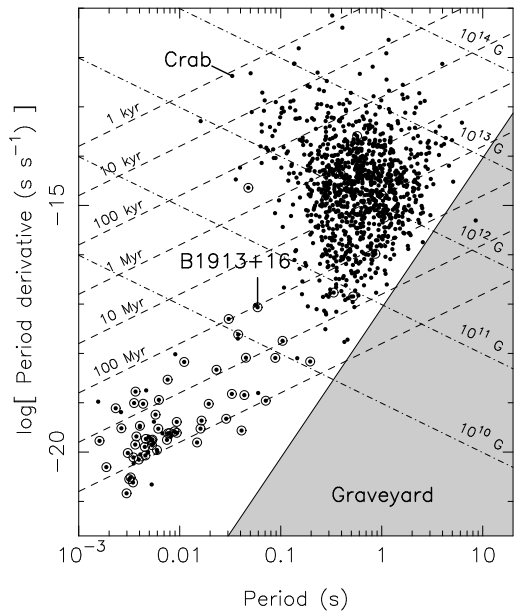}}
\caption[]{The $P-\dot{P}$--diagram for the known pulsar
population. Lines of constant characteristic age and surface
magnetic field are shown. Binary pulsars are marked by
a circle. The solid line represents the pulsar ``death line''
enclosing the ``pulsar graveyard''
where pulsars are expected to switch off radio emission.
}
\label{03fig:ppdot}
\end{figure}

\section{A Pulsar's Life}
\label{03life}

The evolution in pulsar period, $P$,
and slow-down, $\dot{P}$, can be used to describe the
life of a pulsar. This is usually done in a (logarithmic)
$P$-$\dot{P}$-diagram as shown in Fig.~\ref{03fig:ppdot} where we can
draw lines of constant magnetic field (see Eq.~(\ref{03fig:bfield}))
and constant ``characteristic age'' estimated from 
\begin{equation}
\tau = \frac{P}{2\dot{P}} = - \frac{\nu}{2\dot{\nu}},
\label{03age}
\end{equation}
using either period, $P$, or the spin frequency, $\nu$, and
their derivatives in standard units.
This quantity is a valid estimate for the true age under
the assumption that the initial spin period is much smaller than the
present period and that the spin-down is fully determined by magnetic
dipole braking. 
While it had been assumed that pulsars are
born with birth periods similar to that estimated for the Crab pulsar,
$P_0=19$ ms \cite{03lps93}, recent estimates for a growing number
of pulsars suggest a wide range of initial spin periods from
 14 ms  up to 140 ms \cite{03klh+03}.
Pulsars are therefore born in the
upper left area of Fig.~\ref{03fig:ppdot} and move into the central
part where they spent most of their lifetime.

\subsection{Normal Pulsars}

Most known pulsars have spin periods between 0.1 and 1.0 sec with
period derivatives of typically $\dot{P}=10^{-15}$ s s$^{-1}$.
Selection effects are only partly
responsible for the limited number of pulsars known with very long
periods, the longest known period being 8.5 s \cite{03ymj99}.
The dominant effect is due to the ``death'' of pulsars when their
slow-down has reached a critical state. This state seems to
depend on a combination of $P$ and $\dot{P}$ which can
be represented in the $P-\dot{P}$-diagram as a ``pulsar
death-line''. To the 
right and below this line (see Fig.~\ref{03fig:ppdot})
the electric potential above the polar cap may
not be sufficient to produce the particle plasma that
is responsible for the observed radio emission. While this model can
indeed explain the lack of pulsars beyond the death-line, the truth may
be more complicated as the position of the 8.5-sec pulsar deep in the
``pulsar graveyard'' indicates. Nevertheless, it is clear that the
normal life of radio pulsars is limited and that they die eventually
after tens to a hundred million years.

\subsection{Millisecond Pulsars}
\label{03psrwd}

Inspecting the approx.~1600 sources shown in the $P-\dot{P}$-diagram, it
is obvious that the position of a sub-set of about 100 pulsars located
in the lower left part of the diagram cannot be explained by the above
picture of normal pulsar life.  Instead, these pulsars 
simultaneously have small periods (of the order of milliseconds) and small
period derivatives, $\dot{P}\klesssim 10^{-18}$ s s$^{-1}$.  They
appear much older than ordinary pulsars (see Eq.~(\ref{03age}))
and, indeed, these so-called ``millisecond pulsars'' represent the
oldest population of pulsars with ages up to $\sim 10^{10}$ yr. A model for
their evolutionary history was proposed soon after the discovery of
PSR B1937+21 by Backer et al.~in 1982 \cite{03bkh+82}. This first
millisecond pulsar has a period of only 1.56 ms and
remains the pulsar with the shortest period known.

It is suggested that millisecond periods are obtained when mass and
thereby angular momentum is transferred from an evolving binary
companion while it overflows its Roche lobe \cite{03acrs82}.  In this
model, millisecond pulsars are recycled from a dead binary pulsar
via an X-ray binary phase.  This model implies a
number of observational consequences: a) most normal
pulsars do not develop into a millisecond pulsar as they have long lost a
possible companion during their violent birth event;
 b) for surviving binary systems, X-ray binary pulsars
represent the progenitor systems for millisecond
pulsars; c) the final spin period of recycled pulsars depends on
the mass of the binary companion. A more massive companion evolves
faster, limiting the duration of the accretion process;
d) the majority of
millisecond pulsars have low-mass white-dwarf companions as the
remnant of the binary star. These systems evolve from low-mass X-ray
binary systems (LMXBs); e) high-mass X-ray binary systems (HMXBs)
represent the progenitors for double neutron star systems (DNSs). DNSs
are rare since these systems need to survive a second
supernova explosion. The resulting
millisecond pulsar is only mildly recycled with
a period of tens of millisecond.

The properties of millisecond pulsars and X-ray binaries are
consistent with the described picture. For instance, it is striking
that $\sim 80$\% of all millisecond pulsars are in a binary orbit
while this is true for only less than 1\% of the non-recycled
population.  For millisecond pulsars with a low-mass white dwarf
companion the orbit is nearly circular. In case of double neutron star
systems, the orbit is affected by the unpredictable nature of the kick
imparted onto the newly born neutron star in the asymmetric supernova
explosion of the companion. If the system survives, the result is
typically an eccentric orbit with an orbital period of a few hours. As
we will see, both types of system can be used to test different
aspects of gravitational theories.

\section{Pulsars as Clocks}

By
measuring the arrival time of the received ticks of the pulsar clock
very precisely, we can study effects that determine the propagation of
the pulses in four-dimensional space-time. Millisecond pulsars are the
most useful objects for these investigations: their pulse
arrival times can be measured much more accurately than for normal pulsars
(the measurement precision scales essentially with spin frequency) and
their rotation is much smoother, making them intrinsically better
clocks. Specifically, they do not exhibit rotational instabilities
known for normal pulsars, namely ``timing noise'' and
``glitches''. Glitches are associated with young pulsars and they
represent a sudden increase in rotation frequency that is probably
caused by an abrupt change in the internal structure of the neutron
star. The origin of timing noise is much less understood.  It
manifests itself in a quasi-random walk in one or more of the
rotational parameters on timescales of months to years. Again, it
appears mostly for young pulsars and scales with some power of
the period derivative, $\dot{P}$. Hence, millisecond pulsars generally
do not show timing noise, although it has been detected for few
sources such as PSR B1937+21 \cite{03ktr94} albeit on a much smaller
amplitude scale than for normal pulsars. 

\subsection{Time Transfer}

In order to study effects that change the pulse travel time, we first
have to find an expression that describes the pulsar rotation in a
reference frame co-moving with the pulsar. We start by expressing the
spin frequency of the pulsar in a Taylor expansion,
\begin{equation}
\label{03spinevolv}
\nu(t) = \nu_0 +\dot{\nu}_0(t-t_0)+\frac{1}{2}\ddot{\nu}_0(t-t_0)^2
 + ...,
\end{equation}
where $\nu_0$ is the spin frequency at reference time $t_0$,
i.e.~$\nu_0 =\nu(t_0)=1/P_0$ with $P_0$ being the corresponding pulse
period.  While $\nu_0$ and its derivatives refer to values measured
at a certain epoch, 
$\dot{\nu}$, $\ddot{\nu}$ are determined by the physical process
responsible for the pulsar slow-down and should,
in principle, be constant for most
time-spans considered. We expect a relation
\begin{equation}
\label{03nudot}
\dot{\nu} = - \mbox{const.} \; \nu^{n},
\end{equation}
and hence $\ddot{\nu} = - \mbox{const.} \times n 
\times \nu^{n-1} \; \dot{\nu}$
where the ``braking index'', $n$, has a value of $n=3$ for magnetic
dipole braking, relating to Eq.~(\ref{03age}). Measuring $\ddot{\nu}$ 
can yield the braking index via $n = \nu \ddot{\nu}/\dot{\nu}^2$
so that the assumption of dipole braking can be tested. Timing
noise can mimic a significant but time-varying value of $\ddot{\nu}$
that reflects timing noise rather than regular spin-down. In these
cases, derived braking indices are meaningless in terms of 
global spin-down.  For most
millisecond pulsars $\ddot{\nu}$ is too small to be of significance
although some source show a non-zero $\ddot{\nu}$ due to timing noise.

Relating the
spin frequency to the pulse number $N$,  we find
\begin{equation}
\label{03spindown}
N = N_0 + \nu_0(t-t_0) + \frac{1}{2}\dot{\nu}_0(t-t_0)^2
 + \frac{1}{6}\ddot{\nu}_0(t-t_0)^3 + \cdots
\end{equation}
where $N_0$ is the pulse number at the reference epoch $t_0$.  
If $t_0$ coincides with the arrival of a pulse and the
pulsar spin-down is accurately known, the pulses should therefore
appear at integer values of $N$ when observed in 
an inertial reference frame.

Our observing frame is not inertial, as we are using
telescopes that are located on a rotating Earth orbiting the
Sun. Before analysing corresponding TOAs measured with the observatory
clock (``topocentric arrival times''), we need to transfer them to the
centre of mass of the solar system as the best approximation to an
inertial frame available. By using such ``barycentric arrival times'',
we can easily combine transferred topocentric TOAs measured at
different observatories at different times.
The transformation of a topocentric TOA to
a barycentric arrival time, $t_{\rm SSB}$, is given by
\begin{subeqnarray}
t_{\rm SSB} = &  & t_{\rm topo} - t_0 + 
  t_{\rm corr} - D/f^2 \; ,
 \label{03transA} \\
      & + & \Delta_{{\rm Roemer,} \odot} + \Delta_{{\rm Shapiro,}
 \odot} + \Delta_{{\rm Einstein,} \odot} \; , \label{03transB} \\
      & + & \Delta_{\rm Roemer, Bin} +  \Delta_{\rm Shapiro, Bin} +
  \Delta_{\rm Einstein, Bin}. \label{03transC} 
\end{subeqnarray}
We have split the transformation into three lines. The first two lines
apply to every pulsar whilst the third line is only applicable to
binary pulsars. We discuss each term in detail.

\subsubsection{Clock and frequency corrections}

The observatory time is typically maintained by local H-maser clocks
that are compared to UTC (NIST) by the Global Positioning System
(GPS). Offsets are monitored and retroactively applied as clock
corrections, $t_{\rm corr}$, in the off-line analysis which often
uses UTC of BIPM as time standard. Further corrections
take into account that the Earth is not rotating uniformly, so that
leap seconds are occasionally inserted into UTC to keep it close mean
solar time. All leap second are removed from the used UTC time standard to
produce a TOA measured in International Atomic Time (TAI).  The TAI is
maintained as an average of a large number of selected atomic clocks
by the {\em Bureau International des Poids et Mesures} (BIPM), which
also publishes a retroactive uniform atomic time standard known as
Terrestrial Time, TT (formerly known as Terrestrial Dynamical Time,
TDT).  The unit of TT is the SI second and may be regarded as the time
that would be kept by an ideal atomic clock on the geoid with
TT$=$TAI$+$32.184 seconds, where the offset of about 32 s stems from
historic reasons. This time scale should be used in the final analysis
by correcting the initially measured TOAs to TT(BIPM).

As the pulses are delayed due to dispersion in the interstellar
medium, the arrival time depends on the observing frequency, $f$.  The
TOA is therefore corrected for a pulse arrival at an infinitely high
frequency, thereby removing dispersion from the data (last term in
Eq.~\ref{03transA}).  The corresponding Dispersion Measure (DM) is
determined during the discovery of the pulsar and can be measured
accurately by observations at multiple frequencies. For some pulsars,
the dispersion measure is observed to change with time. In order to
avoid time-varying drifts introduced into the TOAs in such cases, the
above term needs to be modified to include time-derivatives of DM,
i.e.  $\dot{\rm DM}$, $\ddot{\rm DM}$ and so on. These can be
determined if monitoring observations at two or more frequencies are
available when they provide an estimate for the change in electron density
along the line-of-sight as a result of `interstellar weather'. For
high-precision timing of millisecond pulsars such multi-frequency
observations are essential.

\subsubsection{Barycentric corrections}

The terms in Eq.~(\ref{03transB}) describe the corrections
necessary to transfer topocentric to barycentric TOAs.

The {\em Roemer delay}, $\Delta_{{\rm Roemer,} \odot}$, is
the classical light-travel time between the
phase centre of the telescope and the solar
system barycentre (SSB). Given a
unit vector, $\vec{{\hat{s}}}$, pointing from the SSB 
to the position of the pulsar and the vector connecting the SSB
to the observatory, $\vec{r}$, we find:
\begin{equation}
\label{03roemer}
\Delta_{{\rm Roemer,} \odot}
= - \frac{1}{c}\; \vec{r} \cdot{\vec{\hat{s}}}
= - \frac{1}{c}\left(
\vec{r}_{\rm SSB} + \vec{r}_{\rm EO} \right) 
\cdot {\vec{\hat{s}}}.
\end{equation}
Here $c$ is the speed of light and we have split $\vec{r}$ into two parts.
The vector, $\vec{r}_{\rm SSB}$, points from the SSB to centre of the
Earth (geocentre). Computation of this vector requires
accurate knowledge of the
locations of all major bodies in the Solar system and uses 
{\em solar system ephemerides} such as the `DE200'
or `DE405' published by {\em Jet Propulsion Laboratory} (JPL)
 \cite{03sta82}. The second vector $\vec{r}_{\rm EO}$,
connects the geocentre with the phase centre of the telescope. In
order to compute this vector accurately, the non-uniform rotation of
the Earth has to be taken into account, so that the correct relative
position of the observatory is derived. This is achieved using 
appropriate UT1 corrections published by
International Earth Rotation Service (IERS).

The {\em Shapiro delay}, $\Delta_{{\rm Shapiro,} \odot}$, is a
relativistic correction that corrects for 
extra delays due to the curvature of
space-time caused by the presence of masses in the solar
system \cite{03sha64}. The delays are largest for a
signal passing the Sun's limb ($\sim 120$ $\mu$s) while
Jupiter can contribute as much as 200 ns. In principle one has to
sum over all bodies in the solar system, yielding 
\begin{equation}
\Delta_{{\rm Shapiro,} \odot} = (1+\hat{\gamma}) \; \sum_i 
\frac{GM_i}{c^3}
\ln \left[ 
\frac{ \vec{\hat{s}} \cdot \vec{r}_i^{E} + r_i^E}{
 \vec{\hat{s}} \cdot \vec{r}_i^P + r_i^P}
\right],
\end{equation}
where $M_i$ is the mass of body $i$, $\vec{r}_i^P$ is the pulsar
position relative to it, and $\vec{r}_i^E$ is the telescope position
relative to that body at the time of closest approach of the photon 
(see \cite{03bh86}). The parameter
$\hat{\gamma}$ is one of the Parameterised-Post-Newtonian
(PPN) parameters that will be discussed in Section~\ref{03ppn}.
It describes how much space-curvature is produced by unit
rest mass and takes the value $\hat{\gamma}=1$ in general relativity
\footnote{Usually, this parameter is denoted without the ``hat''
simply as $\gamma$ (see \cite{03Tur}), but it is common practice in
the study of binary pulsars to use the symbol $\gamma$ to describe the
amount of time dilation and gravitational redshift caused by a 
pulsar companion.}  In
practice, $\hat{\gamma}$ is adopted as unity and only the Sun, and in
some cases Jupiter, are accounted for in this calculation.

The last term in Eq.~(\ref{03transB}), $\Delta_{{\rm Einstein,}
\odot}$, is called {\em Einstein delay} and it describes the combined
effect of time dilation due to the motion of the Earth and 
gravitational redshift caused by the other bodies in the Solar system.
This time varying effect takes into account the variation of an
atomic clock on Earth in the changing gravitational potential
as the Earth follows its elliptical orbit around the
Sun. The delay amounts to an integral of the expression
\cite{03bh86}
\begin{equation}
\frac{d \Delta_{{\mathrm Einstein,} \odot}}{dt}
= \sum_i \frac{G M_i}{c^2 r_i^E} + \frac{v_E^2}{2c^2} -
\mbox{\rm constant},
\end{equation}
where the sum is again over all bodies in the solar system but this
time excluding Earth. The distance $r_i^E$ is again the distance
between Earth and body $i$, while $v_E$ is the velocity of the Earth
relative to the Sun.

\subsubsection{Relative Motion}
\label{03shlovskii}

Equation~\ref{03transA}-\ref{03transC} is sufficient to measure the
clock rate as produced by the pulsar if no further motion or
acceleration between pulsar and SSB occurs. If the pulsar is moving
relative to the SSB, only the transverse component of the velocity,
$v_t$, can be observed from timing. A radial motion is not measurable
practically (though theoretically possible), leaving resulting Doppler
corrections to observed periods, masses etc.~undetermined. The
situation changes if the pulsar has an optically detectable companion
such as a white dwarf for which Doppler shifts can be measured from
optical spectra.  In contrast, a transverse motion will change the vector
$\vec{\hat{s}}$ in Eq.~(\ref{03roemer}), adding a linear
time-dependent term to our transfer equation, and can therefore be
measured as proper motion, $\mu$.

Another effect arising from a transverse motion is the {\em 
Shklovskii}
 {\em effect}, also known in classical astronomy as ``secular
acceleration''. With the pulsar motion, the 
projected distance of the pulsar to
the SSB is increasing, leading to a correction that is quadratic in
time \cite{03bh86},
\begin{equation}
\Delta t_S = \frac{v_{t}^2}{2dc}\; t^2.
\end{equation}
Since this delay
scales with inverse of the distance $d$ to the pulsar,
the correction is usually
too small to be considered. However, it has the effect
that any observed change in a periodicity (i.e.~change in
pulse or orbital period) is increased over the 
intrinsic value by 
\begin{equation}
\frac{\dot{P}}{P} = \frac{1}{c}\;\frac{v_t^2}{d}.
\end{equation}
For millisecond pulsars where $\dot{P}$ is small, a significant
fraction of the observed change in period can be due to the
Shklovskii effect. This effect also needs to be considered, when
studying the decay of an orbital period due to gravitational wave
emission, where the observed value is increased by the Shklovskii
term. 

Similarly, any line-of-sight acceleration $a$ of the pulsar due to an
external gravitational field changes the observed 
period derivative by $\dot{P}/P=a/c$. This effect is commonly
observed for pulsars in globular clusters where the acceleration
through the cluster's gravitational field towards our line of sight
can often be  large enough to reverse the sign of $\dot{P}$. 
As a result, the pulsars appear to be spinning up rather than down!
Such pulsars place useful constraints on the cluster mass distributions 
and the intracluster medium \cite{03phi92b}. 

Finally, a related term that needs to be considered for
nearby pulsars describes an annual parallax 
given by  \cite{03bh86}
\begin{equation}
\Delta t_{\pi} = 
- \frac{1}{2cd} \left( \vec{r}\times \vec{\hat{s}} \right)^2
= \frac{1}{2cd} \left( (\vec{r}\cdot \vec{\hat{s}})^2 - |\vec{r}|^2 \right) .
\end{equation}
In comparison to the more familiar positional parallax, this {\em
timing parallax} corresponds to measuring the time delay due to the
curvature of the emitted wavefronts at different positions of the
Earth orbit. This effect imposes a signal with an amplitude of $l^2
\cos \beta/(2c d)$ where $l$ is the Earth-Sun distance and $\beta$ is
the ecliptic latitude of the pulsar. For a pulsar at $d=1$ kpc, this
delay amounts to only $\klesssim 1.2\mu$s, and hence it is only
measurable for a few millisecond pulsars where it provides a precise
distance estimate. Similarly difficult to measure is the {\em
annual-orbital parallax} for binary pulsars which manifests itself as
a periodic change in the observed projected semi-major axis of the
pulsar's orbit due to viewing the system from slightly different
directions during the Earth's orbit. In contrast, a secular change of
the semi-major axis due to a proper motion of the system on the sky
has been measured for a number of binary millisecond pulsars (see
\cite{03sta03}).

\begin{figure}[h]
\centerline{\includegraphics[width=0.75\textwidth]{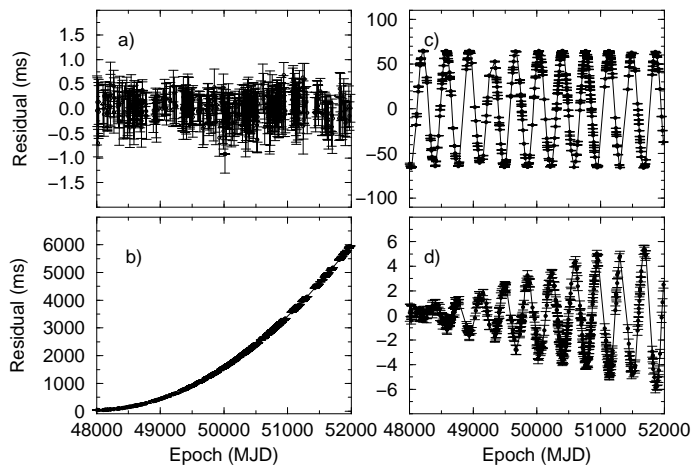}}
\caption[]{Timing residuals for the 1.19-s pulsar B1133+16. A fit 
of a perfect timing model should result in randomly distributed
residuals, shown in a). A parabolic increase in the 
residuals in part b) is obtained if $\dot{P}$
is underestimated, here by 4\%. An offset in position produces
sinusoid residuals shown in part c) where the declination has
an error of 1 arcmin. Part d) demonstrates the effect of 
neglected proper motion, here of $\mu=380$ mas/yr. Note
the different scales on the y-axes.
}
\label{03fig:1133tim}
\end{figure}

\subsection{Pulsar Timing}

The transfer equation Eqs.~(\ref{03transA})--(\ref{03transC}) 
contains a number of parameters
which are not known {\em a priori} (or only with limited precision
after the discovery of a pulsar) and need to be determined precisely
in a least-squares fit analysis of the measured TOAs. These
parameters can be categorised into three groups:

\begin{description}
\item[Astrometric parameters:] The astrometric parameters include
the position of the pulsar, and its proper motion and parallax.
While the position is only known within a telescope beam
after the discovery, the precision can be greatly improved by
timing the pulsar for about a year (full Earth orbit). 
Proper motion and parallax 
only become evident after a longer time-span.
\item[Spin parameters:] These include the rotation frequency of
the pulsar, $\nu$, and its derivatives (\ref{03spinevolv}).
\item[Binary parameters:] For pulsars in a binary orbit, the initial
observations will typically show a periodic variation in observed pulse
period. Five Keplerian parameters then need to be determined:
orbital period, $P_b$; the projected semi-major axis of the
orbit, $x\equiv a \sin i$ where $i$ is the (usually unknown)
inclination angle of the orbit; the orbital eccentricity, $e$; the
longitude of periastron, $\omega$; and the time of periastron
passage, $T_0$. For a number of binary systems this Newtonian
description of the orbit is not sufficient and relativistic
corrections need to be applied, e.g.~$\omega$ is replaced by
$\omega+\dot{\omega}t$. The measurement of the {\em Post-Keplerian}
(PK) {\em parameters} such as $\dot{\omega}$ allows a comparison with
values expected in the framework of specific theories of gravity. We
discuss these aspects further below.
\end{description}
Given a minimal set of starting parameters, a
least-squares fit is needed
to match the measured arrival times to pulse numbers
according to Eq.~(\ref{03spindown}).  The aim is to obtain a
phase-coherent solution that accounts for every single rotation of the
pulsar between two observations. One starts off with a small set of
TOAs that were obtained so closely in time, that the accumulated
uncertainties in the starting parameters do not exceed one pulse
period.  Gradually, the data set is expanded, maintaining coherence in
phase. When successful, post-fit residuals expressed in pulse phase
show a random distribution around zero (see Fig~\ref{03fig:1133tim}).
After starting with fits for only period and pulse reference phase
over some hours and days, longer time spans slowly require fits for
parameters like spin-frequency derivative(s) and position. Incorrect
or incomplete timing models cause systematic structures in the
post-fit residuals identifying the parameter that needs to be included
or adjusted (see Fig.~\ref{03fig:1133tim}).  The precision of the
parameters improves with length of the data span  and the frequency of
observation, but also with orbital coverage in the 
case of binary pulsars.
Sufficient data sets then enable measurements with
amazing precision, e.g.~the period determined for PSR B1937+21 is known
to a relative precision of $10^{-15}$.

Obviously, the above description shows that the process of pulsar
timing is elaborate. Fortunately, sophisticated software packages have
been developed that combine time transfer and the least-squares fit of
the timing model. The three major packages are TEMPO (ATNF/Princeton
University) \cite{03tempo}, PSRTIME (Jodrell Bank Observatory)
\cite{03psrtime}, and TIMAPR (Pushchino Observatory/MPIfR)
\cite{03timapr}.  The most widely used package is TEMPO.

\section{Applications of Pulsars}

Pulsars are unique and versatile objects which can be used to study an
extremely wide range of physical and astrophysical problems.  Beside
testing theories of gravity one can study the Galaxy and the
interstellar medium, stars, binary systems and their evolution, solid
state physics and the interior of neutron stars.  Investigating the
radio emission of pulsars provides insight into plasma physics under
extreme conditions. In the following we will concentrate on the
application of pulsars as clocks, paying in particular attention to
tests of theories of gravity. Some of these tests involve studies of
{\em PPN parameters} and possible related time variation in the
Gravitational Constant, $G$.

\subsection{PPN parameters}
\label{03ppn}

Metric theories of gravity assume 
{\it (i)} the existence of a symmetric metric, {\it (ii)}
that all test bodies follow geodesics of the metric and 
{\it (iii)}
that in local
Lorentz frames the non-gravitational laws of physics are those of
special relativity. Under these conditions we can study metric
theories with the {\em Parameterised Post-Newtonian} (PPN) formalism
by describing deviations from simple Newtonian physics in the
slow-motion and weak-field limit. This is possible in a
theory-independent fashion, such that the only differences in these
theories occur in the numerical coefficients that appear in the
metric, characterised by a set of 10 real-valued 
PPN-Parameters \cite{03nor68b}.  Each of the
parameters can be associated with a specific physical effect, like the
violation of conservation of momentum or equivalence principles, and
certain values are assigned to them in a given theory. Thereby,
comparing measured PPN parameters to their theoretical values can
single out wrong theories in a purely experimental way. A more
complete description of the PPN formalism and the physical meaning of
PPN parameters is presented by 
Turyshev in this volume \cite{03Tur}. A recent 
review was given by Will \cite{03wil01}. Here we summarize the
studies of those PPN parameters that can be constrained by pulsars.
A more detailed account of related pulsar tests is given by
Wex \cite{03wex00,03wex01} and Stairs \cite{03sta03}.

\subsubsection{Violations of the Strong-Equivalence-Principle (SEP)}

The {\em Strong Equivalence Principle} (SEP) is completely embodied
into general relativity, while alternative theories of gravity predict
a violation of some or all aspects of SEP.  The SEP is, according to
its name, stronger than both the {\em Weak Equivalence Principle}
(WEP) and the {\em Einstein Equivalence Principle} (EEP). The WEP
states that all test bodies in an external gravitational field
experience the same acceleration regardless of the mass and
composition. While the WEP is included in all metric theories of
gravity, the EEP goes one step further and also postulates {\em
Lorentz-invariance} and positional invariance.  Lorentz-invariance
means that no preferred frame exists, so the outcome of a local
non-gravitational experiment is independent from the velocity of the
apparatus, while positional invariance renders it unimportant where
this experiment is being performed. The SEP includes both the WEP and
the EEP, but postulates them also for gravitational
experiments. As a consequence, both Lorentz- and positional invariance
should be independent of the gravitational self-energy of the bodies
in the experiment.  Obviously, all bodies involved in terrestrial lab
experiments possess only a negligible fraction of gravitational
self-energy, so that tests of SEP require the involvement of
astronomical objects.

A violation of SEP means that there is a difference between
gravitational mass, $M_g$, and inertial mass, $M_i$.  The difference
can be written as
\begin{equation}
\frac{M_g}{M_i} \equiv 1+\delta(\epsilon) = 1+\eta \epsilon
+{\cal O}(\epsilon^2),
\end{equation}
where $\epsilon$ is the gravitation self-energy in units of $mc^2$ and
$\eta$ is a parameter characterising the violation of SEP. The latter
parameter was introduced by Nordvedt (1968) who suggested studying the
Earth-Moon system to test for violations of SEP. Due to their
different self-energy (Earth: $\epsilon\sim -4.6 \times 10^{-10}$,
Moon: $\epsilon \sim -0.2 \times 10^{-10}$), Earth and Moon would fall
differently in the external gravitational field of the Sun, leading to
a polarization of the Earth-Moon orbit (``Nordvedt-effect'').  
Lunar-laser-ranging experiments
can be used to put tight limits on $\eta$ 
which is a linear combination of PPN parameters representing effects
due to preferred locations, preferred frames and the violation of the
conservation of momentum. However, even in the Earth-Moon case, or
the Solar system in general, the self-energies involved are still
small and do not test the SEP in strong-field regimes where deviations
to higher order terms of $\epsilon$ could be present. It is at that
point where the circular pulsar-white dwarf systems become important.

For neutron stars, $\epsilon \sim 0.15$, which is large, in particular
considering $\epsilon= 0.5$ for a black hole, and much larger than the
self-energy of a white dwarf, $\epsilon\sim 10^{-4}$. Therefore, the
pulsar and white dwarf companion of a binary system should feel a
different acceleration due to the external Galactic gravitational
field if SEP is violated. Similar to the Nordvedt effect, this should
lead to a polarisation of the pulsar-white dwarf orbit
(``gravitational Stark effect'' \cite{03ds91}).  The eccentricity vector
of such a binary system should therefore have two components, one
constant component due to the external acceleration, and another one
that evolves in time following relativistic periastron advance. Since
the present direction of the evolving eccentricity vector is unknown, 
a careful analysis of all relevant systems in a statistical manner
is needed.  Significant contributions to the results are
made by long orbital-period and small-eccentricity
systems, i.e.~where $P_b/e^2$ is large \cite{03wex00}.

\subsubsection{Preferred-Frames \& Conservation Laws}

Some metric theories of gravity violate SEP specifically by predicting
preferred-frame and preferred-location effects.  A preferred universal
rest frame, presumably equivalent with that of the Cosmic Microwave
Background (CMB), may exist if gravity is mediated (in
part) by a long-range vector field. This violation of local
Lorentz-invariance is described by the two PPN parameters $\alpha_1$
and $\alpha_2$. While both parameters can be tightly constrained in
the weak-field limit of the solar system, $\alpha_1$ can also
be studied in the strong-field regime by
analysing the same low-eccentricity pulsar-white dwarf systems 
with a figure-of-merit given by $P_b^{1/3}/e$:
If $\alpha_1$ were different from zero, a binary system moving
with respect to a preferred universal rest frame would again suffer a
long-term change in its orbital eccentricity. In a statistical
analysis similar to that for the study of the gravitational
Stark-effect, the binary 5.3-ms pulsar PSR J1012+5307 is particularly
valuable. It not only has an extremely small orbital
eccentricity, for which only an upper limit of $e<8\times 10^{-7}$
(68\% C.L.) was found from Jodrell Bank and Effelsberg observations
\cite{03lcw+01}, but its
optically detected white dwarf companion also provides full 3-d
velocity information relative to the CMB.
Using this and other systems, Wex \cite{03wex00,03wex01} derives 
$|\alpha_1| < 1.2 \times 10^{-4}$ (95\% C.L.)
which is slightly better than the solar system limit (see \cite{03Tur}).

In cases where theories both violate the Lorentz-invariance and the
conservation of momentum, the equation of motion for a rotating body
in the post-Newtonian limit contains so-called self-acceleration
terms. This self-acceleration of the body's centre depends on the
internal structure of the rotating body and results from the breakdown
in conservation of total momentum.  Another term in the
self-acceleration involves the body's motion relative to a universal
rest frame. Both contributions relate to the PPN parameter $\alpha_3$
that can be tested using pulsars as isolated rotating objects
\cite{03wil93,03bel96}, % will is the book!  
or as bodies in binary systems where both pulsar and companion suffer
self-acceleration, leading to polarized orbits \cite{03bd96}.  The
limits derived by this second method using circular pulsar-white dwarf
systems are much tighter than studying the spin periods of isolated
pulsars, resulting in $|\alpha_3| < 1.5 \times 10^{-19}$ (95\% C.L.)
\cite{03bel96}.  This result for $\alpha_3$ and a limit set on
$(\alpha_3 + \zeta_2) < 4\times 10^{-5}$ \cite{03wil92} constrains the
PPN parameter $\zeta_2$, which describes the non-conservation of
momentum.  Derived from a limit on the second period derivative,
$\ddot{P}$, of PSR B1913+16 and, hence, its acceleration, the
interpretation of this limit may be complicated as non-zero
$\ddot{P}$s could arise from a number of sources such as timing noise
\cite{03sta03}.

\subsubsection{Gravitational Dipole Radiation}

\label{03dipole}

Essentially any metric theory of gravity that embodies
Lorentz-invariance in its field equations predicts gravitational
radiation. However, the details of these predictions may differ in the
speed of the gravity waves, the polarization of the waves and/or the
multi-polarities of the radiation. If a theory satisfies SEP, like
general relativity, gravitational dipole radiation is not expected,
and the quadrupole emission should be the lowest multipole term. This
arises because the dipole moment (centre of mass) of isolated systems
is uniform in time due to the conservation of momentum and because
the inertial mass that determines the dipole moment is the same as the
mass that generates gravitational waves. In alternative theories,
while the {\em inertial} dipole moment may remain uniform, the {\em
gravity wave} dipole moment may not, since in a violation of SEP the
mass generating gravitational waves depends differently on the
internal gravitational binding energy of each body than does the
inertial mass \cite{03wil01}.  
If dipole radiation is predicted, the magnitude of
this effect depends on the difference in gravitational
binding energies, expressed by the difference in coupling
constants to a scalar gravitational field, 
$(\hat{\alpha}_p - \hat{\alpha}_c)$. 
For a white dwarf companion $|\alpha_c| \ll
|\alpha_p|$, so that the strongest emission should occur for
short-orbital period pulsar-white dwarf systems. Again, the binary
pulsar J1012+5307 becomes extremely useful. Given its vanishing
eccentricity, the change in orbital period due to dipole
radiation becomes
\begin{equation}
\dot{P}_b^{\mathrm dipole} \simeq
\frac{4\pi^2 G_\ast}{c^3 P_b} \frac{M_p M_c}{M_p+M_c}
\hat{\alpha_p}^2 + {\cal O}\left( \frac{v^5}{c^5}\right),
\end{equation}
where $G_\ast$ is the ``bare'' gravitational constant.  With the
optically detected companion, the measured radial velocity can be use
to correct for Doppler effects. For this system \cite{03lcw+01}, Wex
\cite{03wex01} derives a limit of $|\hat{\alpha}_p|^2 < 4\times10^{-4}$
(95\% C.L.).

\subsubsection{Time Variability of the Gravitational Constant}

Three different pulsar tests are available to test the time
variability of the gravitational constant, $G$, and to derive an upper
limit for $\dot{G}/G$. A time variability is only allowed if SEP is
violated due to preferred locations in space and time. In the case of
pulsars, a changing $G$ would change the gravitational binding energy
of a neutron star and thereby possibly also its moment of inertia,
which would cause a change in the spin-down behaviour, namely a
contribution to $\dot{P}$. A comparison with observed values leads to
limits of the order of $\dot{G}/G< 10^{-11}$ yr$^{-1}$ \cite{03sta03}.
A slightly more stringent limit can be derived from the effects that a
varying $G$ would have on orbital periods \cite{03dgt88,03arz95}.
 %damour gibbson taylor.  
Both limits are still about an order of magnitude
above the limits set by solar system tests. Moreover, they depend to
some degree on the compactness of the neutron star and its equation of
state, so that they are not truly theory independent \cite{03wil01}. An
interesting alternative test uses the mass determination for neutron
stars \cite{03tho96a}, utilising that the Chandrasekhar mass depends
directly on $G$. Studying the mass of millisecond pulsars as function
of pulsar age, a strong limit of $\dot{G}/G< (-0.6\pm4.2) 10^{-12}$
yr$^{-1}$ (95\% C.L.)  is derived. However, while the mass of neutron
stars can be determined quite accurately in relativistic binaries (see
Section~\ref{03dns}), an age estimation relying on Eq.~(\ref{03age})
can contain considerable uncertainty.

\subsection{Tests using Double Neutron Stars}

Even though in all metric theories matter and non-gravitational fields
respond only to the space-time metric, it is possible that scalar or
vector fields exist in addition to the metric. Damour \&
Esposito-Far{\`e}se developed a framework to study theories at a second
post-Newtonian (2PN) level where gravity is mediated by a tensor field
and one or more scalar field \cite{03de96a}.  These theories are
interesting since scalar partners to gravitons arise naturally in
quantum gravity and unified theories. Damour \& Esposito-Far{\`e}se
show that it is possible to construct corresponding theories where
deviations from general relativity are not visible in the weak field
but only manifest themselves in a ``spontaneous scalarization'' if the
strong field limit is approached. They conclude that current solar
system tests and also upcoming satellite missions will not be able
to replace the strong field tests provided by radio pulsars. Indeed,
they use the DNSs, PSR B1534+12 and PSR
B1913+16, together with PSR B0655+64 and solar system tests, to
significantly constrain the parameters describing the coupling of
matter to the scalar field \cite{03de98}. More stringent limits
have been presented recently using also results for the pulsar-white dwarf
system PSR J1141$-$4565 \cite{03esp04}. A more classical
approach using DNSs for tests of theories of gravity is made
with the measurement of post-Keplerian (PK) parameters as observables.

\label{03dns}
Because of the strong gravitational fields, we expect DNSs to suffer
large relativistic effects.  In this case, we cannot necessarily
assume that we understand the underlying physics, even though general
relativity appears to describe the physics in the solar system to high
precision.  Therefore, a purely theory-independent approach like the
PPN approximation is difficult to realize.  Instead, one can only use
an existing theory of gravity and check if the observations are
consistently described by the measured Keplerian and PK parameters.
In each theory, for point masses with negligible spin contributions,
the PK parameters should only be functions of the a priori unknown
pulsar and companion mass, $M_p$ and $M_c$, and the easily measurable
Keplerian parameters. With the two masses as the only free parameters,
an observation of two PK parameters will already
determine the masses uniquely in the framework of the given
theory. The measurement of a third or more PK parameters then provides
a consistency check. In general relativity, the five most important PK
parameters are given to lowest Post-Newtonian order
by (e.g.~\cite{03wex01}):
\begin{eqnarray}
\dot{\omega} &=& 3 T_\odot^{2/3} \; \left( \frac{P_b}{2\pi} \right)^{-5/3} \;
               \frac{1}{1-e^2} \; (M_p + M_c)^{2/3}, \label{03omegadot}\\
\gamma  &=& T_\odot^{2/3}  \; \left( \frac{P_b}{2\pi} \right)^{1/3} \;
              e\frac{M_c(M_p+2M_c)}{(M_p+M_c)^{4/3}}, \\
\dot{P}_b &=& -\frac{192\pi}{5} T_\odot^{5/3} \; \left( \frac{P_b}{2\pi} \right)^{-5/3} \;
               \frac{\left(1 +\frac{73}{24}e^2 + \frac{37}{96}e^4 \right)}{(1-e^2)^{7/2}} \; 
               \frac{M_pM_c}{(M_p + M_c)^{1/3}}, \\
r &=& T_\odot M_c, \\
s &=& T_\odot^{-1/3} \; \left( \frac{P_b}{2\pi} \right)^{-2/3} \; x \;
              \frac{(M_p+M_c)^{2/3}}{M_c},
\end{eqnarray}
where $P_b$ is the period and $e$ the eccentricity of the binary
orbit. The masses $M_p$ and $M_c$ of pulsar and companion,
respectively, are expressed in solar masses ($M_\odot$).  We define
the constant $T_\odot=GM_\odot/c^3=4.925490947 \mu$s where $G$ denotes
the Newtonian constant of gravity and $c$ the speed of light. The
first PK parameter, $\dot{\omega}$, is the easiest to measure and
describes the relativistic advance of periastron.
According to Eq.~(\ref{03omegadot}) it provides 
an immediate measurement of the total mass of the system,
$(M_p+M_c)$. The parameter
$\gamma$ denotes the amplitude of delays in arrival times caused by
the varying effects of the gravitational redshift and time dilation
(second order Doppler) as the pulsar moves in its elliptical orbit at
varying distances from the companion and with varying speeds.  The
decay of the orbit due to gravitational wave damping is expressed by
the change in orbital period, $\dot{P}_b$. The other two parameters,
$r$ and $s$, are related to the Shapiro delay caused by the
gravitational field of the companion. These parameters are only
measurable, depending on timing precision, if the orbit is seen nearly
edge-on.

\begin{figure}[h]
\begin{minipage}{6.cm}
\centerline{\includegraphics[width=0.8\textwidth]{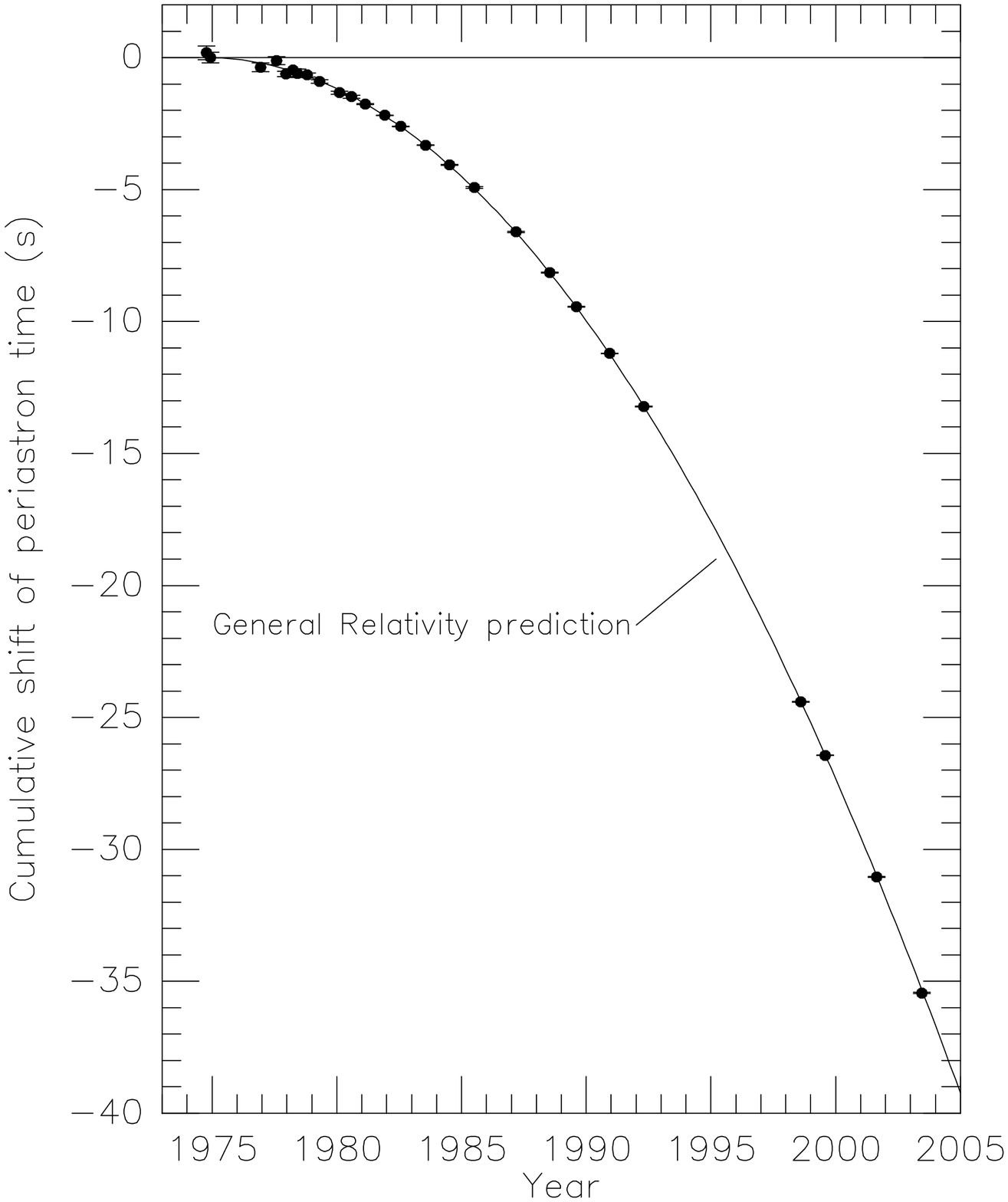}}
\end{minipage}
\hspace{0.cm}
\begin{minipage}{6.cm}
\centerline{\includegraphics[width=1.\textwidth]{030737m1m2.ps}}

\end{minipage}
\bigskip
\caption[]{(left)
Shift in the periastron passage of the DNS PSR B1913+16 plotted as a
function of time, resulting from orbital energy loss
due to the emission of gravitational radiation. The agreement between
the data, now spanning almost 30 yr, and the predicted curve due to
gravitational quadrupole wave emission is now better than
0.5\%. Figure provided by Joel Weisberg and Joe Taylor.
(right) ``Mass-mass'' diagram showing the
observational constraints on the masses of the neutron stars in the
double-pulsar system J0737--3039.  The shaded regions are those which
are excluded by the Keplerian mass functions of the two
pulsars. Further constraints are shown as pairs of lines enclosing
permitted regions as predicted by general relativity: (a) the
measurement of $\dot{\omega}$ gives the total system mass
$m_A+m_B=2.59$ M$_{\odot}$; (b) the measurement of the mass ratio
$R=m_A/m_B=1.07$; (c) the measurement of the gravitational
redshift/time dilation parameter $\gamma$; (d) the measurement of the
two Shapiro delay parameters $r$ and $s$. Inset is an enlarged view of
the small square encompassing the intersection of the three tightest
constraints, representing the area allowed by general relativity
and the present measurements.
}
\label{03fig:dns}
\end{figure}

Until very recently, 
only two DNSs had more than two PK parameters determined,
the 59-ms pulsar B1913+16 and the 38-ms PSR B1534+12.  For PSR
B1913+16 with an eccentric ($e=0.61$) 7.8-hr orbit, the PK parameters
$\dot{\omega}$, $\gamma$ and $\dot{P}_b$ are measured very
precisely. Correcting the observed $\dot{P}_b$ value for effects of
relative motion (see Section~\ref{03shlovskii}), the measured value is in
excellent agreement with the prediction of general relativity for
quadrupole emission (see Fig.~\ref{03fig:dns}). This result
demonstrates impressively that general relativity provides a
self-consistent and accurate description of the system which can be
described as orbiting point masses, i.e.~the structure of the neutron
stars does not influence their orbital motion as expected from
SEP. The precision of this test is limited by our knowledge of the
Galactic gravitational potential and the corresponding 
correction to $\dot{P}_b$.  The timing results for PSR B1913+16
provide us with the most precise measurements of neutron star masses
so far, i.e.~$M_p=(1.4408\pm0.0003)M_\odot$ and $M_c=(1.3873\pm0.0003)
M_\odot$ \cite{03wt03}. It is worth pointing out that these values
include the unknown Doppler factor.

The 10-hr orbit of the second DNS PSR B1534+14 ($e=0.27$) is observed
under fortunate circumstances, it is seen nearly edge-on. Thereby, in
addition to the three PK parameters observed for PSR B1913+16, the
Shapiro-delay parameters $r$ and $s$ can be measured, enabling
non-radiative aspects of gravitational theories to be tested, as
$\dot{P}_b$ is not necessarily needed. In fact, the observed value of
$\dot{P}_b$ seems to be heavily influenced by Shklovskii-terms, so that
the corresponding line fails to meet the others in a $M_p$-$M_c$
diagram. However, assuming that general relativity is the correct
theory of gravitation, the deviation from the predicted value 
and the measured proper motion, $\mu$, can 
be used to compute the necessary correction and hence the distance to
the pulsar, $d=1.02\pm0.05$ kpc \cite{03sttw02}.

\subsection{Tests Using Profile Structure Data}

In addition to the use of pulsars as clocks, strong gravity effects
can also be tested using pulse structure data, namely the effects of
``geodetic precession'' in the DNSs PSR B1913+16 and PSR B1534+14. In
both cases, the pulsar spin axis appears to be misaligned with the
orbital angular momentum vector. In such a case, general relativity
predicts a relativistic spin-orbit coupling, analogous to spin-orbit
coupling in atomic physics.  The pulsar spin precesses about the total
angular momentum, changing the relative orientation of the pulsar
towards Earth. As a result, the angle between the pulsar spin axis and
our line-of-sight changes with time, so that different portions of the
emission beam are observed \cite{03dr74}.  Consequently, changes in
the measured pulse profile and its polarization are expected.  In
extreme cases, the precession may even move the beam out of our
line-of-sight and the pulsar may disappear as predicted for PSR
B1913+16 for the year 2025
\cite{03kra98}.  See the review by Kramer \cite{03kra01} for
a detailed description of this effect and its observations.

\subsection{Recent Discoveries}
\label{03future}
The tremendous success of recent surveys, in particular those using
the Parkes telescope (e.g.~\cite{03mlc+01}), has lead not only to the
discovery of more than 700 new pulsars, but also to some very exciting
new binary systems.  Until recently, only five DNSs were known. This
situation has changed and now eight systems can be studied. The most
recent addition is a DNS discovered in the Parkes Multibeam (PM)
Survey, PSR J1756$-$2251 (Faulkner et al., in prep.). 
This system shows similarities with PSR
B1913+16 as its orbital period is somewhat less than 8 hours although
its eccentricity is smaller ($e=0.18$). 

The recent discoveries benefit from larger available computing power
which enables so-called ``acceleration searches'' for fast orbiting
binary pulsars. Such techniques try to correct for the usually made
assumption that the pulse period remains constant during the
observations.  For compact binary systems, this assumption is 
violated due to large Doppler shifts in period, resulting in much
reduced sensitivity in standard Fourier searches. The employment
of acceleration codes has therefore lead to a number of new binary
pulsars with
short orbital periods. Another example is PSR J1744$-$3922 (Faulkner
et el., in prep.). This 172-ms pulsar is in an almost circular
4.6-hr orbit and hence only the second long-period pulsar in such a
short orbit. The other such pulsar is PSR J1141$-$6545 which is a
393-ms PM pulsar in an eccentric
4.5-hr orbit \cite{03klm+00a}. Both pulsars appear
to have a white dwarf companion, but while PSR J1141$-$6545's
companion is heavy ($M_c\sim 1M_\odot$), the new pulsar's companion is
probably much lighter ($M_c\ge0.08M_\odot$). Whilst these are indeed
exciting discoveries, the most stunning success is clearly the
recent discovery of the first double-pulsar system, J0737$-$3039.

\section{The Double-Pulsar}

The 22.8-ms pulsar J0737-3039 was discovered in April 2003
\cite{03bdp+03}. It was soon found
to be a member of the most extreme relativistic binary system ever
discovered: its short orbital period ($P_b = 2.4$ hrs) is combined
with a remarkable high value of periastron advance
($\dot{\omega}=16.88\pm 0.09 \deg$/yr, i.e.~four times larger than for
PSR B1913+16!) and a short coalescing time ($\sim 85$ Myr).  The
latter time-scale boosts the hopes for detecting a merger of two
neutron stars with first-generation ground-based gravitational wave
detectors by about an order of magnitude compared to previous
estimates based on only the DNSs B1534+12 and B1913+16
\cite{03bdp+03}. Consequently,
during the lecture I had presented this pulsar already as the most
beautiful laboratory for testing general relativity found so far,
pointing also out that with a geodetic precession period of only 70 yr
future studies should reveal interesting and exciting results. But
little did we know then which surprise was still waiting for us.

In October 2003, our team detected radio pulses from the second
neutron star when data sets covering the full orbital period were
analysed \cite{03lbk+04}. The reason why signals from the 2.8-s pulsar
companion (now called PSR J0737$-$3039B, hereafter ``B'') to the
millisecond pulsar (now called PSR J0737$-$3039A, hereafter ``A'') had
not been found earlier, became clear when it was realized that B was
only visible clearly for two short parts of the orbits.  For the
remainder of the orbit, the pulsar B is extremely weak and only
detectable with the most sensitive equipment. The detection of a young
pulsar-companion B clearly confirmed the evolution scenario presented
in Section~\ref{03life} and made this already exciting system
sensational, providing a truly unique testbed for
relativistic gravity.

Indeed, we have now measured A's $\dot{\omega}$ and $\gamma$ and we have
also detected the Shapiro delay in the pulse arrival times of A due to
the gravitational field of B, providing a precise measurement of the
orbital inclination of $\sin i = 0.9995(^{+4}_{-32})$. Obviously, as
another strike of luck, we are observing the system almost completely
edge-on which allows us to also probe pulsar magnetospheres for the
very first time by a background beacon. The measurements already
provide four measured PK parameters, resulting in a $m_A-m_B$ plot
shown in Fig.~\ref{03fig:dns}. The orbital decay due to gravitational
wave emission is already visible in the data, but the uncertainties
are yet too large to provide a useful constraint.  However, in
addition to tests with these PK parameters, the detection of B as a
pulsar opens up opportunities that go well beyond what has been
possible so far. With a measurement of the projected semi-major axes
of the orbits of both A and B, we obtain a precise measurement of the
mass ratio, $R(m_{\rm{A}}, m_{\rm{B}}) \equiv m_{\rm{A}}/m_{\rm{B}} =
x_{\rm{B}}/x_{\rm{A}}$, providing a further constraint displayed in
Fig.~\ref{03fig:dns}.  For every realistic theory of gravity, we can
expect the mass ratio, $R$, to follow this simple relation
\cite{03dt92}. Most importantly, the $R$-line is not only
theory-independent, but also independent of strong-field (self-field)
effects which is not the case for PK-parameters.  This provides a
stringent and new constraint for tests of gravitational theories as
any intersection of the PK-parameters {\em must} be located on the
$R$-line. At the same time, it provides us already with very accurate
mass measurements for the neutron stars, $M_A=(1.337\pm0.005)M_\odot$
and $M_B=(1.250\pm0.005)M_\odot$, respectively, making B the
least-massive neutron star ever observed.

The equations for the PK parameters given in Section~\ref{03dns} are
all given to lowest Post-Newtonian order. However, higher-order
corrections may become important if relativistic effects are large and
timing precision is sufficiently high. Whilst this has not been the
case in the past, the double pulsar system may allow measurements of
these effects in the future \cite{03lbk+04}. One such effect involves
the prediction by general relativity that, in contrast to Newtonian
physics, the neutron stars' spins affect their orbital motion via
spin-orbit coupling. This effect would be visible clearest as a
contribution to the observed $\dot{\omega}$ in a secular \cite{03bo75}
and periodic fashion \cite{03wex95}.  For the J0737$-$3039 system, the
expected contribution is about an order of magnitude larger than for
PSR B1913+16, i.e.~$2\times 10^{-4}$ deg yr$^{-1}$ (for A, assuming a
geometry as determined for PSR B1913+16 \cite{03kra98}). As the exact
value depends on the pulsars' moment of inertia, a potential
measurement of this effect allows the moment of inertia of a neutron
star to be determined for the first time \cite{03ds88}.  Obviously,
the double pulsar system offers improved but also new tests of general
relativity. The current data already indicate an agreement of the
observed with the expected Shapiro parameter of $s_{\rm obs}/s_{\rm
exp} = 1.00007\pm0.00220$ (Kramer et al.~in prep.) where the
uncertainties are likely to decrease.

\section{Conclusions \& Outlook}

Millisecond pulsars find a wide range of applications, in particular
for precise tests of theories of gravity.  After the discovery of
pulsars thereby marked the beginning of a new era in fundamental
physics, pulsars discovered and observed with the future
Square-Kilometer-Array (SKA) will further transform our understanding
of gravitational physics.  The SKA's sensitivity will discover the
majority of pulsars in the Galaxy, almost certainly providing the
discovery of the first pulsar-black hole system. For tests of general
relativity such a system would have with a discriminating power that
surpasses all its present and foreseeable competitors \cite{03de98}.
In particular, we could directly test BH properties as predicted by
general relativity, such as the Cosmic Censorship Conjecture or the
``no-hair'' theorem.  Moreover, the pulsars discovered with the SKA
would act as arms of a
huge gravitational wave detector enabling the study of a possible
gravitational wave background in a frequency range that is
inaccessible to LIGO or even LISA. Clearly, the SKA will provide yet
another leap in our understanding and application of pulsars.

\label{03_}

%%%{\include{kramer}}
%%%


\begin{thebibliography}{88.}


\bibitem{03wil01} C.~M. Will, {Living Rev. Relativity} {\bf 4}, 4. [Online article]: cited on 1 Oct 2003, http://www.livingreviews.org/lrr-2001-4 (2001)

\bibitem{03Tur} S. G. Turyshev et al.: In: {\it Astrophysics, Clocks and Fundamental Constants}, ed. by S. G. Karshenboim and E. Peik, Lecture Notes in Physics Vol. 648 (Springer, Berlin, Heidelberg 2004)

\bibitem{03wex01} N. Wex, In: {\it Gyros, Clocks, Interferometers...: Testing Relativistic
  Gravity in Space}, eds C. L{\"a}mmerzahl,   C. W. F. Everitt, \&  F.~W. Hehl, (Springer, 2001)

\bibitem{03sta03} I.~H. Stairs,  {Living Rev. Relativity} {\bf 6}, 5. [Online article]: cited on 1 Oct 2003,
  http://www.livingreviews.org/lrr-2003-5 (2003)

\bibitem{03lor01} D.~R. Lorimer, {Living Rev. Relativity} {\bf 4},  5. [Online article]: cited on 1 Oct 2003,
  http://www.livingreviews.org/lrr-2001-5 (2001)

\bibitem{03tc99} S.~E. Thorsett \& D. Chakrabarty,  {Ap. J.} {\bf 512}, 288 (1999)

\bibitem{03ov39} J.~R. Oppenheimer \& G. Volkoff, {Phys. Rev.} {\bf 55}, 374 (1939)

\bibitem{03zp98} V.~E. Zavlin and G.~G. Pavlov, A\&A {\bf 329}, 583 (1998)

\bibitem{03msk+03} M.~A. McLaughlin, I.~H. Stairs, V.~M. Kaspi et al., Ap. J. {\bf 591}, L135 (2003)

\bibitem{03bclm03} G.~F. Bignami,  P.~A. Caraveo, A.~D. Luca \&  S. Mereghetti, Nature {\bf 423}, 725 (2003)

\bibitem{03hkwe03}T.~H. Hankins, J.~S. Kern, J.~C. Weatherall \& J.~A. Eilek, 
 {Nature} {\bf 422}, 141 (2003)

\bibitem{03kxj+97} M. Kramer, K.~M. Xilouris,~A. Jessner, et al.,
A\&A, {\bf 322}, 846 (1997)

\bibitem{03lps93} A.~G. Lyne, R.~S. Pritchard \& F.~G. Smith, 
 MNRAS {\bf 265}, 1003 (1993)

\bibitem{03klh+03} M. Kramer, A.~G. Lyne, G. Hobbs, et al.,
 Ap.J. {\bf 593}, L31 (2003)

\bibitem{03ymj99} M.~D. Young, R.~N. Manchester \& S. {Johnston},  Nature {\bf 400}, 848 (1999) 

\bibitem{03bkh+82} D.~C. Backer, S.~R. Kulkarni, C. Heiles, M.~M. Davis \&  W.~M.Goss, Nature {\bf 300}, 615 (1982)   

\bibitem{03acrs82} M.~A. Alpar, A.~F. Cheng, M.~A. Ruderman \& J. Shaham, Nature {\bf 300}, 728 (1982)

\bibitem{03ktr94} V.~M. Kaspi, J.~H. Taylor \& M. Ryba, Ap. J. {\bf 428}, 713 (1994)

\bibitem{03sta82} E.~M. Standish, A\&A {\bf 114}, 297 (1982)

\bibitem{03sha64} I.~I. Shapiro, Phys. Rev. Lett. {\bf 13}, 789 (1964) 

\bibitem{03bh86} D.~C. Backer \& R.~W. Hellings, Ann. Rev. Astr. Ap. {\bf 24}, 537 (1986)

\bibitem{03phi92b}  E.~S. Phinney, Philos. Trans. Roy. Soc. London A{\bf 341}, 39
(1992)

\bibitem{03tempo} http://pulsar.princeton.edu/tempo/
  
\bibitem{03psrtime} http://www.jb.man.ac.uk/research/pulsar/observing/progs/progs.html

\bibitem{03timapr} http://www.mpifr-bonn.mpg.de/div/pulsar/former/olegd/soft.html

\bibitem{03nor68b} K. Nordtvedt, Phys. Rev. {\bf 170}, 1186 (1968)

\bibitem{03wex00} N. Wex, In: {\em Pulsar Astronomy - 2000 and Beyond, {IAU} Colloquium 177},
  eds M.~Kramer, N.~Wex \& R.~Wielebinski, R., ASP Conf.~Series Vol.~202
  (PASP, San Francisco 2000), p.~113

\bibitem{03ds91} T. Damour \& G. Sch\"afer, Phys. Rev. Lett. {\bf 66}, 2549 (1991)
 
\bibitem{03lcw+01} C. Lange, F. Camilo, N. Wex et al., MNRAS {\bf 326}, 274 (2001) 

\bibitem{03wil93} C.~M. Will, {\em Theory and Experiment in Gravitational Physics}, 
  (Cambridge University Press, Cambridge 1993)  

\bibitem{03bel96} J.~F. Bell, Ap. J. {\bf 462}, 287 (1996)

\bibitem{03bd96}Bell, J.~F. \& Damour, T., Class.~Quantum Grav.,
 {\bf 13}, 3121 (1996)

\bibitem{03wil92} C.~M. Will, Ap. J. {\bf 393}, L59 (1992)
 
\bibitem{03dgt88} T. Damour, G.~W.  Gibbons \&  J.~H. Taylor, Phys. Rev. Lett. {\bf 61}, 1151 (1988)
 
\bibitem{03arz95} Z. Arzoumanian, {\em PhD thesis}, Princeton University (1995)

\bibitem{03tho96a} S.~E. Thorsett, Phys. Rev. Lett. {\bf 77}, 1432 (1996)

\bibitem{03wt03} J.~M. Weisberg \& J.~H. Taylor, In: {\em Radio Pulsars}, eds M.~Bailes, D.J.~Nice \& S.E.~Thorsett,
 ASP Conf.~Series Vol.~302 (PASP, San Francisco 2003), p.~93

\bibitem{03de96a} T. Damour \& G. Esposito-Far\`ese, Phys. Rev. D{\bf 53}, 5541 (1996)

\bibitem{03de98} T. Damour \& G. Esposito-Far\`ese, Phys. Rev. D{\bf 58}, 042001 (1998)

\bibitem{03esp04}
{Esposito-Far\`ese}, G.
contribution to 10th Marcel Grossmann meeting, gr-qc/0402007 (2004)

\bibitem{03sttw02} I.~H. Stairs, S.~E. Thorsett, J.~H. Taylor \& A. Wolszczan, Ap. J. {\bf 581}, 501 (2002)

\bibitem{03dr74} T. Damour \& R. Ruffini, Academie des Sciences Paris Comptes Rendus Ser.\,Scie.\,Math. 
{\bf 279}, 971 (1974)

\bibitem{03kra98} M. Kramer, Ap. J. {\bf 509}, 856 (1998)

\bibitem{03kra01} M. Kramer, In: {\em The Ninth Marcel Grossmann Meeting},
 eds V.G.~Gurzadyan, R.T.~Jantzen \& R.~Ruffini (World Scientific, Singapore 2002) p.~219

\bibitem{03mlc+01} R.~N. Manchester, A.~G. Lyne, F. Camilo, et al., MNRAS {\bf 328}, 17 (2001)
 
\bibitem{03klm+00a}  V.~M. Kaspi, A.~G. Lyne, R.~N. Manchester, et al., Ap. J. {\bf 543}, 321 (2000)

\bibitem{03bdp+03} 
{Burgay}, M., {D'Amico}, N., Possenti, et al., Nature, {\bf 426}, 531 
(2003)

\bibitem{03lbk+04}
Lyne, A.~G., {Burgay}, M., Kramer, M., et al., Science, {\bf 303}, 1153
(2004)

\bibitem{03dt92} Damour, T. \& Taylor, J.~H., Phys. Rev. D, 
{\bf 45}, 1840 (1992)

\bibitem{03bo75} {Barker}, B.~M. \& {O'Connell}, R.~F., 
Phys. Rev. D, {\bf 12}, 329 (1975)

\bibitem{03wex95}
Wex, N., 1995, 
\newblock {\it Class. Quantum Grav.}, {\bf 12}, 983 (1995)

\bibitem{03ds88} Damour, T. \& Sch{\"a}fer, G., Nuovo Cim., 
{\bf 101}, 127 (1988)


\end{thebibliography}
\end{document}